\documentclass[%
 reprint,
superscriptaddress,
 amsmath,amssymb,
aps,
]{revtex4-2}
\usepackage{mhchem}
\usepackage{siunitx}
\usepackage{graphicx}
\usepackage{dcolumn}
\usepackage{bm}
\usepackage{hyperref}
\usepackage[mathlines]{lineno}

\begin{document}

\title{Vortex-driven periodic and aperiodic magnetoresistance oscillations in cuprates}

\author{Changshuai Lan}\affiliation{Ministry of Education Key Laboratory of Fundamental Physical Quantities Measurement and Hubei Key Laboratory of Gravitation and Quantum Physics, National Precise Gravity Measurement Facility and School of Physics, Huazhong University of Science and Technology, Wuhan 430074, People’s Republic of China}
\author{Chuanwen Zhao}\affiliation{Ministry of Education Key Laboratory of Fundamental Physical Quantities Measurement and Hubei Key Laboratory of Gravitation and Quantum Physics, National Precise Gravity Measurement Facility and School of Physics, Huazhong University of Science and Technology, Wuhan 430074, People’s Republic of China}
\author{Xin Yi}\affiliation{Ministry of Education Key Laboratory of Fundamental Physical Quantities Measurement and Hubei Key Laboratory of Gravitation and Quantum Physics, National Precise Gravity Measurement Facility and School of Physics, Huazhong University of Science and Technology, Wuhan 430074, People’s Republic of China}
\author{Qiao Chen}\affiliation{Ministry of Education Key Laboratory of Fundamental Physical Quantities Measurement and Hubei Key Laboratory of Gravitation and Quantum Physics, National Precise Gravity Measurement Facility and School of Physics, Huazhong University of Science and Technology, Wuhan 430074, People’s Republic of China}
\author{Xinming Zhao}\affiliation{Ministry of Education Key Laboratory of Fundamental Physical Quantities Measurement and Hubei Key Laboratory of Gravitation and Quantum Physics, National Precise Gravity Measurement Facility and School of Physics, Huazhong University of Science and Technology, Wuhan 430074, People’s Republic of China}
\author{Dong Wu}\affiliation{Beijing Academy of Quantum Information Sciences, Beijing 100193, People’s Republic of China}
\author{Chengyu Yan}\email{chengyu_yan@hust.edu.cn}\affiliation{Ministry of Education Key Laboratory of Fundamental Physical Quantities Measurement and Hubei Key Laboratory of Gravitation and Quantum Physics, National Precise Gravity Measurement Facility and School of Physics, Huazhong University of Science and Technology, Wuhan 430074, People’s Republic of China}\affiliation{Institute for Quantum Science and Engineering, Huazhong University of Science and Technology, Wuhan 430074, People’s Republic of China}
\author{Shun Wang}\email{shun@hust.edu.cn}\affiliation{Ministry of Education Key Laboratory of Fundamental Physical Quantities Measurement and Hubei Key Laboratory of Gravitation and Quantum Physics, National Precise Gravity Measurement Facility and School of Physics, Huazhong University of Science and Technology, Wuhan 430074, People’s Republic of China}\affiliation{Institute for Quantum Science and Engineering, Huazhong University of Science and Technology, Wuhan 430074, People’s Republic of China}

\date{\today}

\begin{abstract}
 The study of the interaction between superconductivity and charge ordering is helpful to resolve the pairing mechanism in high-temperature superconductors.  Recently, several  resistance oscillations studies trigger the speculation that a long-range charge ordering, with an enormous mesh size of several tens of nanometer, can possibly emerge in underdoped high $T_{c}$ superconductor. However, spectroscopy studies have not traced this kind of long-range charge ordering. Here, we clarify the disagreement between the transport and spectroscopy studies on the mysterious long-range charge ordering by investigating the magneto-oscillations in underdoped ${\mathrm{Bi}}_{2}{\mathrm{Sr}}_{2}{\mathrm{CaCu}}_{2}{\mathrm{O}}_{8+\ensuremath{\delta}}$ flakes. Inspired by the observation that the oscillations evolve from a periodic to an aperiodic one with decreasing doping level, we conclude that the magneto-oscillations can be generated by the interaction between vortices and superconducting loops that enclose randomly distributed underdoped puddles while an assumption of long-range charge ordering is not necessary. 
\end{abstract}

\maketitle

High-temperature superconductors, such as ${\mathrm{Bi}}_{2}{\mathrm{Sr}}_{2}{\mathrm{CaCu}}_{2}{\mathrm{O}}_{8+\ensuremath{\delta}}$ (BSCCO), have been a versatile platform for exploring emergent phenomena\cite{RevModPhys.78.17,SongMa-339,KeimerKivelson-449,PhysRevX.10.011056}. In these materials, the anisotropy in the order parameter usually leads to competition and coexistence between superconductivity and other low-temperature phases\cite{HamidianEdkins-785,ChenHashimoto-1495,YangLiu-2842}, which in turn hinders a comprehensive understanding of superconductivity in high-temperature superconductors. The competition is even more prominent because of enhanced fluctuations when the materials are exfoliated to thin flakes\cite{14,17,LiaoZhu-244,PhysRevLett.122.247001,ZouHao-2024,GuWan-299}.

Among all the low temperature phases coexisting with superconductivity, charge ordering is particularly important since it is widely observed in cuprates and constitutes a considerable portion of the phase diagram\cite{RevModPhys.78.17,HamidianEdkins-785}. Recently, several transport studies reported on puzzling long-range charge ordering in two dimensional superconductors including cuprates\cite{8,10,13}, with a mesh size as large as $\sim$50 nm, as evidenced by the emergence of magneto-oscillations. This type of long-range charge ordering can potentially be the key to understand effects such as locally fragile superconductivity\cite{HsuBerben-2754}. However, such a long-range charge ordering has not been directly observed by resonant X-ray scattering (RXS) or scanning tunneling microscopy(STM)\cite{31,32}. Hence, it is an outstanding question to unveil the relation between the magneto-oscillations and the proposed charge ordering.

Here, we report the magnetotransport properties of exfoliated BSCCO flakes at different doping levels. Magneto-oscillations are observed in the vortex liquid state. It is particularly striking to note that the magneto-oscillations become more pronounced while they are more aperiodic. By analyzing the correlation between the phase diagram of underdoped BSCCO and the emergence of magneto-oscillations, we suggest that the interaction between vortices and superconducting loops that enclose randomly distributed underdoped puddles is sufficient to result in the oscillations. The evolution of the oscillations from periodic to aperiodic can be qualitatively explained by the competition between the puddle size distribution and vortex-loop interaction energy costs. Our results highlight that the magneto-oscillations do not necessarily signify the occurrence of long-range charge ordering, and hence shed light on the difference between transport and spectroscopy studies. 

 We modulate the doping level of BSCCO flakes by vacuum annealing them at 300 K\cite{14} so that the oxygen is naturally released from BSCCO (see Supplemental Material for basic characterization\cite{SM}). The results reported in the main text are rendered from sample S1 unless stated otherwise, measurements on other samples can be found in the Supplemental Material\cite{SM}. Fig. \ref{1}(a) shows the temperature-dependent resistances at different doping levels (controlled by the annealing time). Two transition temperatures can be recognized. The first one, the critical temperature $T_{c,max}$ for the trivial superconductivity, occurs around 88 K. $T_{c,max}$ is insensitive to the doping level. Moving to lower temperature, the resistance first enhances and then drops to zero again at the second transition temperature $T_{c}$. The nonmonotonic evolution is different from the smooth two-step transition in vaccum annealed BSCCO monolayer or lithium intercalated flake\cite{14,LiaoZhu-244,13}, which arises from the reentrant superconductivity widely observed in cuprates due to inhomogeneous doping\cite{16,17,PhysRevLett.108.067004}. $T_{c}$ drops with the decreasing doping level (or increasing annealing time). The exact doping level can be extracted from $T_{c}$ by the empirical relation: $T_{
c}=T_{c,max}\left[{\mathrm{1}}-{\mathrm{82.6}}\left(p-{\mathrm{0.16}}\right) \right]^{2}$. Fig. \ref{1}(b) shows the annealing time dependent doping level spanning from $p=0.16$ to 0.051. Besides, the size of the inhomogeneous region is relatively small since the doping level has little impact on both $T_{c,max}$ and normal state resistance.

\begin{figure}[tb]
	\centering
	\includegraphics[width=8.5cm]{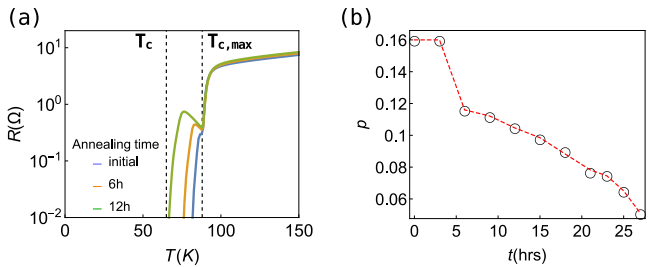}
	\caption{Doping level modulation. (a) Temperature dependent resistance with different doping levels realized by vacuum annealing. The black dashed lines are two critical temperatures. (b) Doping level as a function of annealing time.}
	\label{1}
\end{figure}

\begin{figure}[tb]
	\centering
	\includegraphics[width=8.5cm]{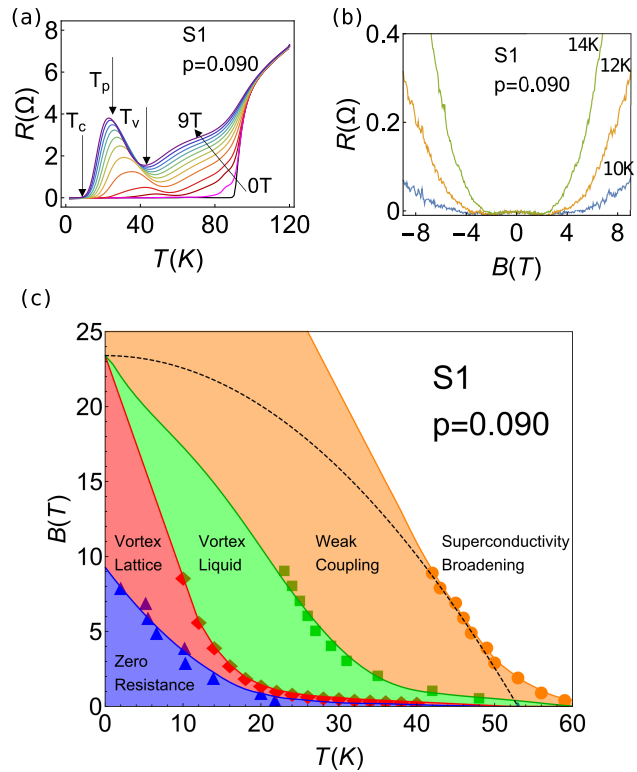}
	\caption{Magnetoresistance of sample S1. (a) Resistance of an underdoped state (p=0.090) as a function of temperature under different transverse magnetic fields varying from 0T to 9 T (0, 0.1, 0.5 T for the lowest three traces and then increments by 1T steps for other traces). (b) Resistance as a function of magnetic field at different temperatures. A full plot that covers the entire resistance range can be found in the Supplemental Material\cite{SM}. (c) The main plot shows the B-T phase diagram. The blue triangles show the critical temperature ${T}_{c}$ for reentrant superconductivity. The red diamonds show the vortex melting field ${B}_{vs}$ defined by the point where the magnetoreistance reaching $1\%$ of the normal state resistance. The green squares show the local maximum point of the temperature dependent resistance ${T}_{p}$. The orange circles show the local minimum point of the temperature dependent resistance ${T}_{v}$. Different phases are shown with different colors. The boundaries between different phases are generated by smooth connecting and the linear extrapolating the characteristic data points. The black dashed curve in the weak coupling regime shows the upper critical field ${B}_{c2}$ based on GL theory. }
	\label{2}
\end{figure}

\begin{figure*}[tb]
	\centering
	\includegraphics[width=17.2cm]{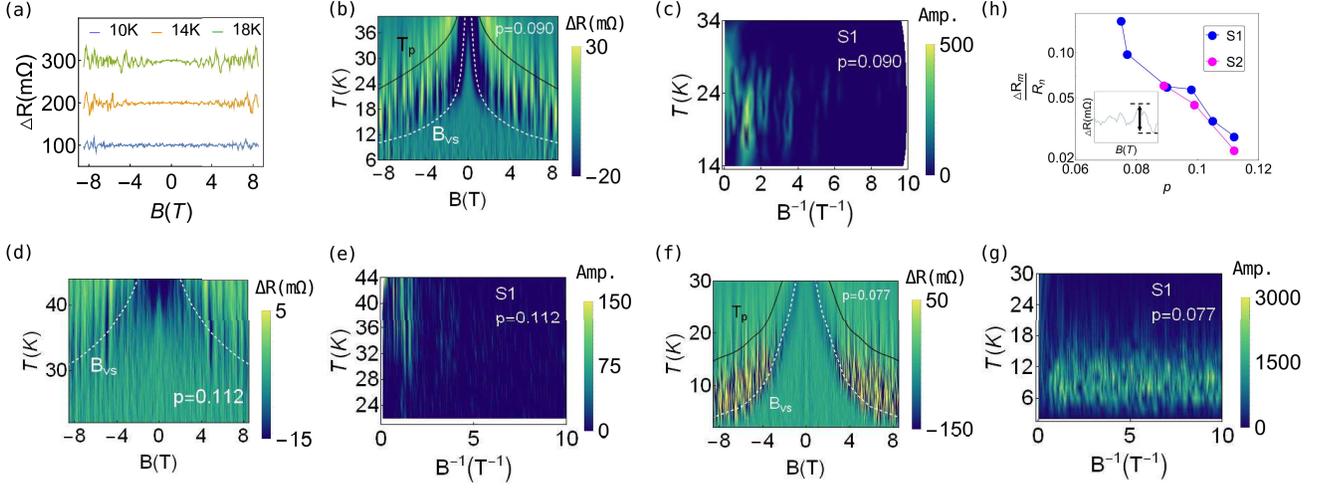}
	\caption{Characteristics of magneto-oscillations. (a) The same data as Fig.\ref{2} after background substraction. The curves are vertically offset for clarity. (b) Color plot of $\Delta R\left(B\right)$ at $p=0.090$.  Oscillations occur between ${T}_{p}$ (black solid trace) and ${B}_{vs}$ (white dashed trace). (c) FFT of $\Delta R\left(B\right)$ in (b). (d)-(g) show the magneto-oscillation and the corresponding FFT data for $p=0.112$ and $p=0.077$, respectively. (h) The oscillation amplitude after normalization against the normal state resistance increases monotonically with reduced doping level. The raw amplitude follow the same trend. The inset shows the definition of $\Delta R_{m}\left(T\right)$. }
	\label{3}
\end{figure*}

  Next, we establish a comprehensive phase diagram at each doping level by studying both temperature and magnetic field dependence. Figure. \ref{2} summarizes the result with a doping level $p=0.090$ (data for other doping levels can be found in Supplemental Material\cite{SM}). First of all, we measure the temperature dependence at fixed magnetic fields, as shown in Fig. \ref{2}(a). It is clear that $T_{c}$ for reentrant superconductivity, $T_{p}$ for resistance maximum, and $T_{v}$ for resistance minimum shift toward lower temperatures with increasing magnetic field, accompanied by the usual broadening of superconductivity transition at $T_{c,max}$. These observations are results of the vortex moving\cite{18}. Then, we sweep the magnetic at given temperatures far below $T_{c,max}$, as depicted in Fig. \ref{2}(b). A finite resistance sets in at a small magnetic field, which reflects a transition from vortex lattice to vortex liquid\cite{19}. The vortex melting field $B_{vs}$ is defined by the point at which the magnetoresistance reaches $1\%$ of the normal state resistance $R_{n}$ (see Supplemental Material for the detail\cite{SM}.). From data presented in Fig. \ref{2}(a)$\&$(b), one can extract the phase diagram Fig. \ref{2}(c). For a given field, the vortex generation and mobilization process drive the system from the reentrant superconductivity state (the zero resistance state marked in the phase diagram) to the weak coupling state. In this process, the aforementioned vortex dynamics cause the nonmonotonic resistance evolution. It is especially necessary to emphasize that the resistance soars in the vortex liquid regime where the vortices are mobile, and we will demonstrate that it is in this very regime where pronounced magneto-oscillations can be observed.

\begin{figure*}[tb]	
\centering
	\includegraphics[width=17.2cm]{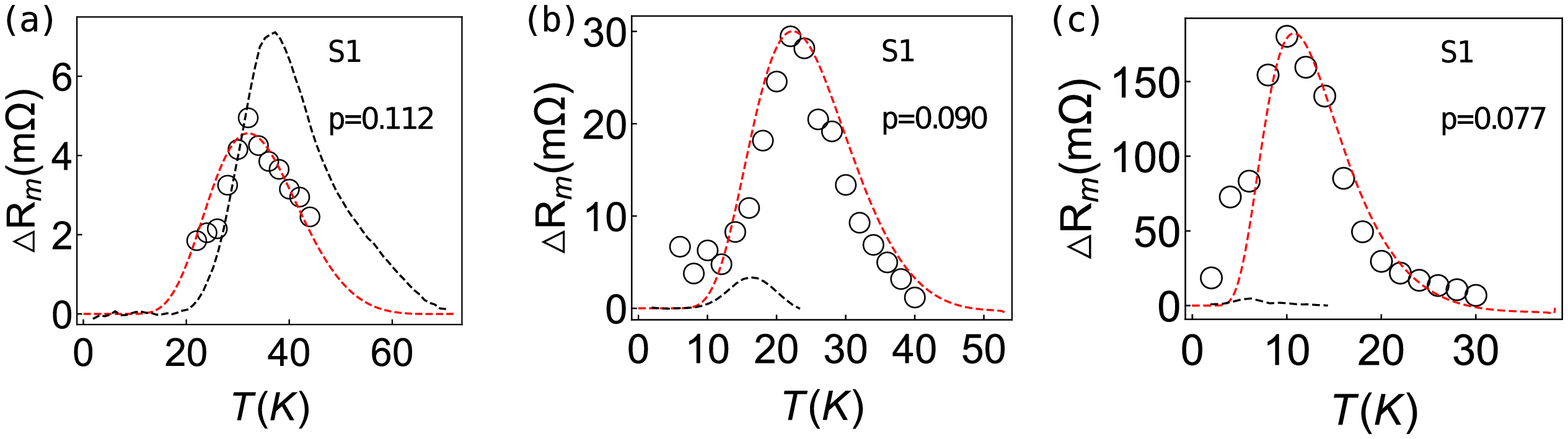}
	\caption{The oscillation amplitude as a function of temperature. The black dashed curve shows the amplitude expected from the Little-Parks effect. The red dashed line represents the calculated results from the vortex effect, see Eq.\ref{eq1}}
	\label{4}
\end{figure*}

 A zoom-in of the data in Fig. \ref{2}(b), where $p=0.090$, uncovers a quasi-periodic oscillation superposed on a smooth background. Figure. \ref{3}(a) shows the oscillations at different temperatures after background subtraction (see Supplemental Material for the detail\cite{SM}.), whereas Fig. \ref{3}(b) summarizes the data in a color plot. The oscillations always occur above the ${B}_{vs}$ and are most promient in the region enclosed by the boundary ${B}_{vs}$ and ${T}_{p}$, i.e., the vortex liquid regime. The fast Fourier transformation (FFT) reveals that the periodicity of the oscillations is well-defined within specific temperature window and becomes more scattered outside the window. We mark $p=0.090$ as a transition regime to reflect the evolution in the periodicity. Similar oscillations in the vortex liquid state can be observed at other doping levels. Increasing the doping level will drive the system into the ordered regime where the oscillations are highly periodic in the temperature range of interest, as shown in Fig. \ref{3}(d)$\&$(e) for $p=0.112$. On the other hand, lowering the doping level makes the system more disordered, where it is difficult to identify the dominant periodicity of the oscillations, as depicted in Fig. \ref{3}(f)$\&$(g) for $p=0.077$. It is also important to highlight that the amplitude of the oscillations increases by orders when the system becomes more disordered with doping level reducing, see Fig. \ref{3}(h).

 These observations are noticeably different from the previous experiments\cite{13,22,YangLiu-2842,25}. For instance, ref\cite{13} reported the oscillations in the vortex lattice state of lightly underdoped BSCCO and found that the oscillation periodicity remained the same regardless of the doping level.
 
  \begin{figure}[htbp]
	\centering
	\includegraphics[width=8.3cm]{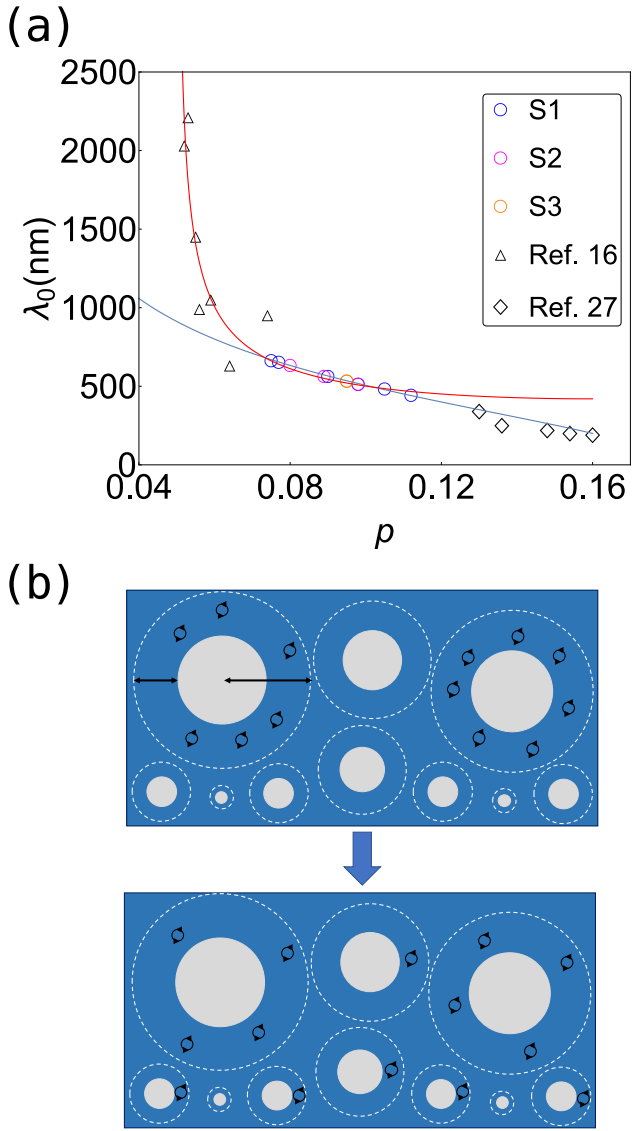}
	\caption{Proposed origin of the oscillation. (a) Doping dependence of penetration depth $\lambda_{0}$ gathered from sample S1-3 and literature\cite{13,34}. The solid curves are fitting results with different models. (Details can be found in Supplemental Material.\cite{SM}) (b) Schematic drawing of interaction between the vortices (black circles) and superconducting loops enclosing underdoped puddles (gray circles). The white dashed circles are superconducting loops. The black circles with arrow schematically illustrate the distribution of vortices. The vortices can drift in and out of the superconducting loop by costing energy $E_l\propto\frac{1}{\lambda_{0}^{2}\cdot{A}}$. For higher doping level (upper panel), the vortices mainly interact with large loops; on the other hand, both large and small loops contribute at lower doping level (lower panel). The seemingly more vortices in the large loops only reflects the fact it is easier for the vortices to interact with the large loops. }
	\label{5}
\end{figure}
 
Regardless of the detail of the experimental setup, previous works come to an agreement that the oscillations are closely related to the vortex dynamics. This can be verified by studying the amplitude of the oscillations\cite{13,22}. In our experiment, the oscillation does not arise from the Little-Parks (LP) effect\cite{21}, which fails to capture the temperature dependence of oscillation amplitude ${\Delta}{R_m}$, as shown in Fig.\ref{4}. Instead, the oscillations can be interpreted by the vortex effect, formulated as\cite{22}:  
\begin{equation}
	\Delta R = R_{n}(\frac{2E_{l}}{k_{B}T})^{2}\frac{I_{1}\left\lbrack \frac{E_{l} + E_{v}}{{2k}_{B}T} \right\rbrack}{\left( I_{0}\left\lbrack \frac{E_{l} + E_{v}}{{2k}_{B}T} \right\rbrack \right)^{3}}
	\label{eq1}
\end{equation}
Here $R_{n}$ is the normal state resistance. $I_{0}$ and $I_{1}$ are the zero- and first-order modified Bessel function of the first kind. The two energy quantities are: $E_{l}={{\Phi}_{0}^{2}}/8{\pi}{{\mu}_{0}}{\Lambda}{(T)}{\cdot}W/A$ and $E_{v}={{\Phi}_{0}^{2}}/2{\pi}{{\mu}_{0}}{\Lambda}{(T)}{\cdot}{\ln{(2W/{\pi}{\xi}{(T)})}}$, where ${{\mu}_{0}}$ is the vacuum permeability, $A$ measures the averaged size of the underdoped regions, $W$ is the averaged width of the constrictions formed between the underdoped regions. The Pearl penetration ${\Lambda}{(T)}=2{\lambda}{(T)}^{2}/d$, with $d$=1.5 nm representing the thickness of monolayer. The temperature-dependent penetration depth and coherence length are: ${\lambda}{(T)}={{\lambda}_{0}/{\sqrt{1-{(T/{T_{c}})}^{2}}}}$ and ${\xi}{(T)}=0.74{{\xi}_{0}/{\sqrt{1-(T/{T_{c}})}}}$. ${{\xi}_{0}}$, ${{T}_{c}}$ and $A$ can be determined experimentally, whereas ${{\lambda}_{0}}$, ${{R}_{n}}$ and $W$ are fitting parameters. The penetration length $\lambda_{0}$ shows a negative correlation with the doping level, as summarized in Fig.\ref{5}(a). Our results supplement the
doping level not covered by previous experiments\cite{13,34}. We can extract a doping-level-independent mean free length $l$=1.1 nm from $\lambda_{0}$(Details can be found in Supplemental Material\cite{SM}). The small mean free path is consistent with the decay length of vortex cores in BSCCO measured by scanning tunneling spectroscopy \cite{PhysRevLett.85.1536}, which further confirms the role of vortex dynamics in the magneto-resistance.

Let us now try to address the origin of the oscillations. It is generally believed that long-range charge ordering and vortex dynamics are the two gradients for the occurrence of magneto-oscillations in unpatterned cuprates. The long-range charge ordering itself is speculated to be a superposition of CDW of different types\cite{8,10,13}. The hypothesized CDW-superposition scenario is not favored in our experiment. Previous experiments suggest that the disorder in cuprates will only enhance the amplitude of CDW or fully destroy it rather than induce a change in periodicity\cite{PhysRevLett.113.107002,PhysRevB.92.224504}. Hence, it is unlikely to cause an enhancement in the magneto-oscillations while the oscillations become more aperiodic. 

Instead, we can interpret our results qualitatively by the interaction between vortices and superconducting loops that enclose randomly distributed underdoped puddles. As shown in  Fig.\ref{5}, the underdoped region forms puddles located randomly in space\cite{14}, the size probability distribution follows a rapid decay with respect to the puddle size $A$\cite{CampiBianconi-2917}. The vortices drift in/out of the superconducting loops that enclose single or multiple puddles, by costing energy $E_l\propto\frac{1}{\lambda_{0}^{2}\cdot{A}}$\cite{PhysRevB.68.214505,PhysRevB.69.064516}. This interaction results in magneto-oscillations\cite{22,SM}. There are more small loops, but it is more energetically costly for the vortex to interact with them; on the contrary, it is energetically favorable for the vortex to interact with large loops, but the large loops are rare cases. The periodicity of the oscillations is determined by the competition between the puddle size distribution and the energy cost. When the doping level is high ($\lambda_0$ is small), the vortex cannot interact with the small loops due to the over-large energy cost $E_l$, instead, it primarily interacts with the rare large loops, so that oscillations have a well-defined periodicity corresponding to the size of large loops. When the doping level is low ($\lambda_0$ is large), both the small and large loops can contribute to the oscillations thanks to the reduced $E_l$, and they can be equally favored at a suitable doping level, so that the distribution of periodicity broadens. The amplitude is mainly determined by $\lambda_0$, it is amplified with decreasing doping levels. Therefore, we have observed periodic oscillations with smaller amplitude when the system is ordered (high doping) and aperiodic oscillations with larger amplitude when the system is disordered (low doping).

In conclusion, we studied the magnetoresistance in underdoped BSCCO flakes with doping levels tuned by vacuum annealing. We observed pronounced magneto-oscillations in the vortex liquid state as a consequence of the interaction between vortices and superconducting loops that enclose randomly distributed underdoped puddles. By analyzing the oscillation periodicity at each doping level, we suggest that the oscillations do not necessarily implicate the emergence of charge ordering. Our results clarify the disagreement between the spectroscopy and transport studies on the mysterious long-range charge ordering in cuprates. 
 
\begin{acknowledgments}
We acknowledge the support from the National Natural Science Foundation of China (12074134, 12204184).
\end{acknowledgments}

\bibliography{ref}

\end{document}